\documentclass{article}
\usepackage{spconf,amsmath,graphicx}
\usepackage{multirow}
\usepackage{float}

\usepackage{enumitem}
\setlist{nosep, leftmargin=14pt}
\usepackage{url}
\usepackage{mwe} 
\usepackage{hyperref}
\usepackage{makecell}
\usepackage{amssymb,amsmath,amsthm,enumitem}

\newcommand\figref{Fig.~\ref}
\newcommand\tabref{Table~\ref}

\makeatletter
\newcommand*\titleheader[1]{\gdef\@titleheader{#1}}
\AtBeginDocument{%
  \let\st@red@title\@title
  \def\@title{%
    \bgroup\normalfont\large\centering\@titleheader\par\egroup
    \vskip.5em\st@red@title}
}
\makeatother
\title{Crop and Couple: Cardiac Image Segmentation using Interlinked Specialist Networks}
\titleheader{This paper has been accepted for ISBI-2024}

\name{Abbas Khan$^{1,2}$, Muhammad Asad$^{3}$, Martin Benning$^{2,4}$, Caroline Roney$^{2,5}$, Gregory Slabaugh$^{1,2}$}
\address{$^{1}$School of Electronic Engineering and Computer Science, Queen Mary University of London, UK\\
$^{2}$Queen Mary's Digital Environment Research Institute (DERI), London, UK\\
$^{3}$School of Biomedical Engineering and Imaging Sciences, King’s College London, UK\\
$^{4}$Department of Computer Science, University College London, UK\\
$^{5}$School of Engineering and Materials Science, Queen Mary University of London, UK}

\begin{document}
%
\maketitle
\begin{abstract}

Diagnosis of cardiovascular disease using automated methods often relies on the critical task of cardiac image segmentation. We propose a novel strategy that performs segmentation using specialist networks that focus on a single anatomy (left ventricle, right ventricle, or myocardium). Given an input long-axis cardiac MR image, our method performs a ternary segmentation in the first stage to identify these anatomical regions, followed by cropping the original image to focus subsequent processing on the anatomical regions. The specialist networks are coupled through an attention mechanism that performs cross-attention to interlink features from different anatomies, serving as a soft relative shape prior. Central to our approach is an additive attention block (E-2A block), which is used throughout our architecture thanks to its efficiency. 
The source code is available at \footnote[1]{https://github.com/kabbas570/CroCNet}. 

\end{abstract}

\section{Introduction}\label{sec:intro}
Cardiovascular diseases are the leading cause of death globally, claiming the lives of approximately 17.9 million people annually, according to the World Health Organization 
\footnote[1]{https://www.who.int/health-topics/cardiovascular-diseases\#tab=tab\_1}.
In the pursuit of early detection and treatment planning, Magnetic Resonance Imaging (MRI) is an important modality, as it is non-invasive and provides detailed images of the heart anatomy. This enables clinical experts to assess cardiac function, identify structural irregularities, and plan effective treatment strategies \cite{cohn2003screening}. Deep learning is a powerful tool in cardiac image computing 
\cite{chen2020deep} to assist in clinical workflows.

Deep learning-based methods have paved the way for automated partitioning of semantically meaningful regions of interest in medical images \cite{ronneberger2015u}.  In the context of cardiac images, these regions include the left ventricle (LV), right ventricle (RV), and Myocardium (MYO), among others. Automatic segmentation provides diagnostic markers, facilitating the identification and assessment of diverse pathologies within the respective cardiac regions.
\begin{figure*}[ht!]
\centering
\includegraphics[scale=0.6]{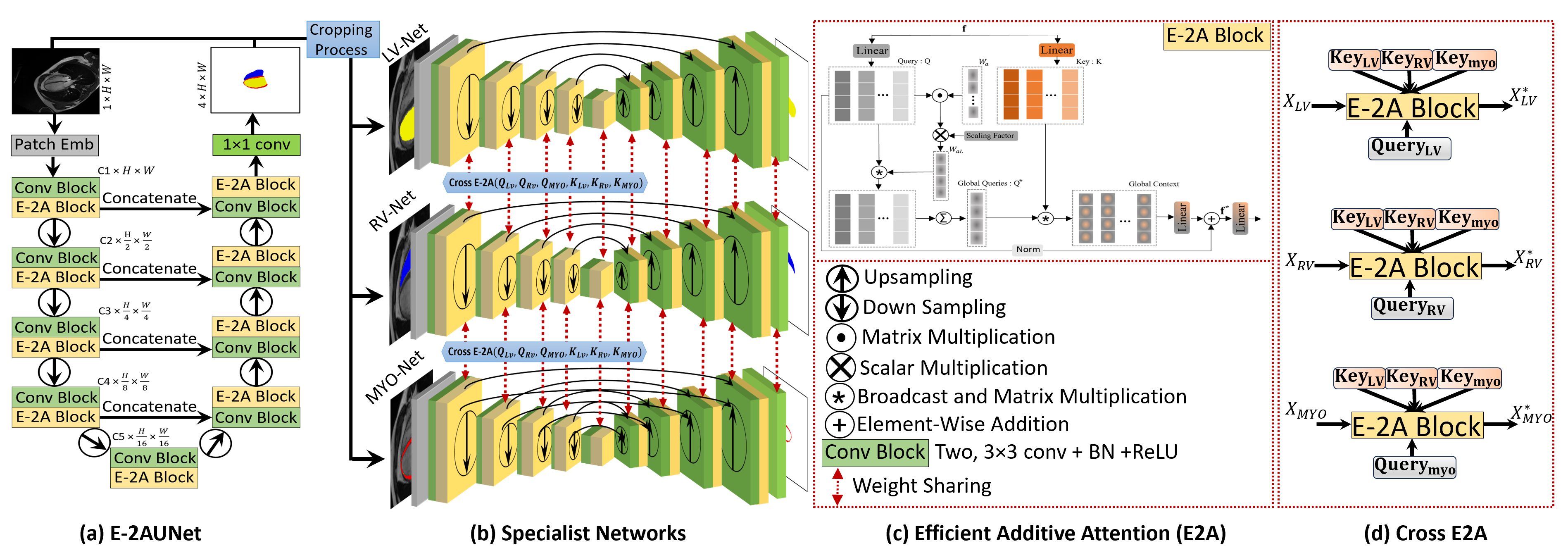}
\caption{Overview of the proposed CroCNet pipeline.  (a) E-2AUNet architecture for initial segmentation in the first stage. 
The cropped image and initial binary segmentations are fed into (b) specialist networks (LV-Net, RV-Net, and MYO-Net) to refine the predictions for each cardiac region. We leverage (c) efficient additive attention, including (d) a cross-E2A block that implements cross-attention to intermingle features between the specialist networks.} 
\label{fig1}
\end{figure*}
Prior work on cardiac image segmentation includes traditional models like active shape models (ASM) \cite{de2003adapting} and 
atlas-based methods \cite{kiricsli2010evaluation}. Recent studies have explored deep learning models such as convolutional neural networks (CNNs) \cite{baccouch2023comparative}, and self-attention architectures like TransBridge \cite{deng2021transbridge} and Transfusion \cite{liu2022transfusion}. While these methods have shown great promise for cardiac image segmentation, there still remains room for improvement.

In this paper, we propose a novel pipeline called  \textbf{Cro}p and \textbf{C}ouple (\emph{CroCNet}), which implements a two-stage cardiac image segmentation pipeline. The first stage consists of an \textbf{E}fficient \textbf{A}dditive \textbf{A}ttention-based UNet (\emph{E-2AUNet}) architecture 
that provides an initial segmentation and is used to crop the image.  The second stage of CroCNet utilizes weight-sharing and cross-attention mechanisms to correct the errors in the initial segmentation. Both stages leverage the recently proposed additive attention of \cite{shaker2023swiftformer}, a highly efficient approach to computing attention. Our contributions are as follows:

\begin{enumerate}
    \item We propose CroCNet, a novel two-stage architecture that computes a first segmentation used to identify anatomies and perform cropping on the original image. The cropped data is fed into a second stage consisting of coupled specialist networks which perform binary segmentation. 
    \item For the first stage of CroCNet, we propose E-2AUNet, a novel hybrid encoder-decoder architecture that modifies a UNet with E-2A blocks.  
    \item In the second stage, we implement specialist networks coupled through efficient additive cross-attention, which acts as a soft shape prior. 
    \item Our results set a new state-of-the-art on the M\&Ms-2 dataset, outperforming the recently proposed TransFusion \cite{liu2022transfusion} method with a large improvement of 3.19\% Dice score and 4.67mm Hausdorff distance on average while only requiring a single view input.
\end{enumerate}
\section{Related Work}
U-Net \cite{ronneberger2015u} pioneered deep learning architectures for medical image segmentation by proposing an encode-decoder architecture. 
This architecture has inspired the development of numerous extensions, such as ResUNet \cite{diakogiannis2020resunet}, U-Net++ \cite{zhou2018unet++}, and InfoTrans \cite{li2021right}. However, these methods struggle to effectively learn long-range semantic information interaction due to the nature of localized convolution operations.\\
Motivated by the vision transformer \cite{carion2020end}, several methods have been proposed to incorporate multi-head self-attention (MHSA) 
for medical image segmentation. TransUNet \cite{chen2021transunet} adopts a U-shaped architecture by utilizing the transformer layers in the encoder blocks. UTNet \cite{gao2021utnet} integrates MHSA in both encoder-decoder blocks and proposes a revised MHSA to reduce computational complexity from $O({n}^2$) to $O(n)$. Multi-Compound Transformer (MCTrans) \cite{ji2021multi} utilizes multi-scale features extracted by CNNs and adds the contextual information in skip connections via Self-Attention and Cross-Attention block. TransFusion \cite{liu2022transfusion} improves \cite{gao2021utnet} by using Divergent Fusion Attention to model different views' features in the encoder.
\section{Proposed Method}\label{sec: Proposed Method}
Our proposed approach, \emph{CroCNet}, implements a two-stage method, shown in \figref{fig1}, E-2AUNet and specialist networks.

\subsection{E-2AUNet}
The segmentation network employed in the first stage of \emph{CroCNet} is the E-2AUNet, depicted in \figref{fig1}(a). For an input image of size 1$\times$H$\times$W, the patch embedding layer produces C1 feature maps followed by a convolution block, consisting of two consecutive 3$\times$3 convolutions with ReLU and batch normalization. The E-2A block is then utilized to capture global information, with this process iterated four times. E-2AUNet can efficiently capture both local and global contexts using convolutional and E-2A blocks. Between the two stages, a downsampling layer reduces spatial dimensions and doubles the channel dimension. The number of feature maps at each stage is increased as C1:32 $\rightarrow$ C2:64 $\rightarrow$ C3:128 $\rightarrow$ C4:256 $\rightarrow$ C5:512. On the decoder side, the convolution block and E-2A block pattern are repeated four times, and an upsampling layer decreases channel numbers while enlarging spatial features. Skip connections are incorporated from the encoder to the decoder for improved gradient flow and connectivity between feature maps. The final segmentation map, representing LV, RV, MYO, and background, is generated using a 1$\times$1 convolution layer with four kernels and sigmoid activation.

\subsection{Efficient-Additive Attention}
The E-2A serves as a fundamental component of the proposed E-2AUNet, illustrated in \figref{fig1}(c). This building block addresses computational inefficiencies associated with self-attention by replacing quadratic complexity with element-wise operations \cite{shaker2023swiftformer}, thereby eliminating the need for key-value interactions. The feature matrix (\emph{\textbf{f}}) is projected into a query (\emph{Q}) and key (\emph{K}). The parameter vector ${\textbf{w}_a}$ is then applied to the query matrix, producing learnable 
attention weights ${\textbf{w}_{aL}}$ , followed by a scaling operation. Subsequently, these weights are multiplied by the original query matrix, resulting in a global attention query vector and summation to get a single global query ${\mathbf{Q}}^{*}$. 
\setlength{\belowdisplayskip}{1pt}
\setlength{\abovedisplayskip}{1pt}
\begin{equation}\label{eq1}
{\mathbf{Q}}^{*}  =  \sum_{n=1}^{n} \Bigl(\frac{\mathbf{Q} \cdot \textbf{w}_{a}}{\sqrt{d}}\Bigl).\mathbf{Q}_{n}
\end{equation}
The global context is established through the interaction of the key matrix with the global query. A linear transformation layer is employed for the global context to capture query-key interactions and hidden token representations.
\begin{equation}\label{eq2}
{\emph{\textbf{f}}}^{*}  =  norm({\mathbf{Q}}) + {Linear}({\mathbf{Q}} * \mathbf{K})
\end{equation}

\subsection{Second Stage Specialist Networks}
CroCNet's second stage consists of specialist networks and cropping, as depicted in \figref{fig1}(b).  Serving as a refinement step, this stage aims to rectify any inaccuracies present in the predictions made during the first stage. We provide details of each component in this stage below.

\begin{figure}[hbt!]
\centering
\includegraphics[scale=0.52]{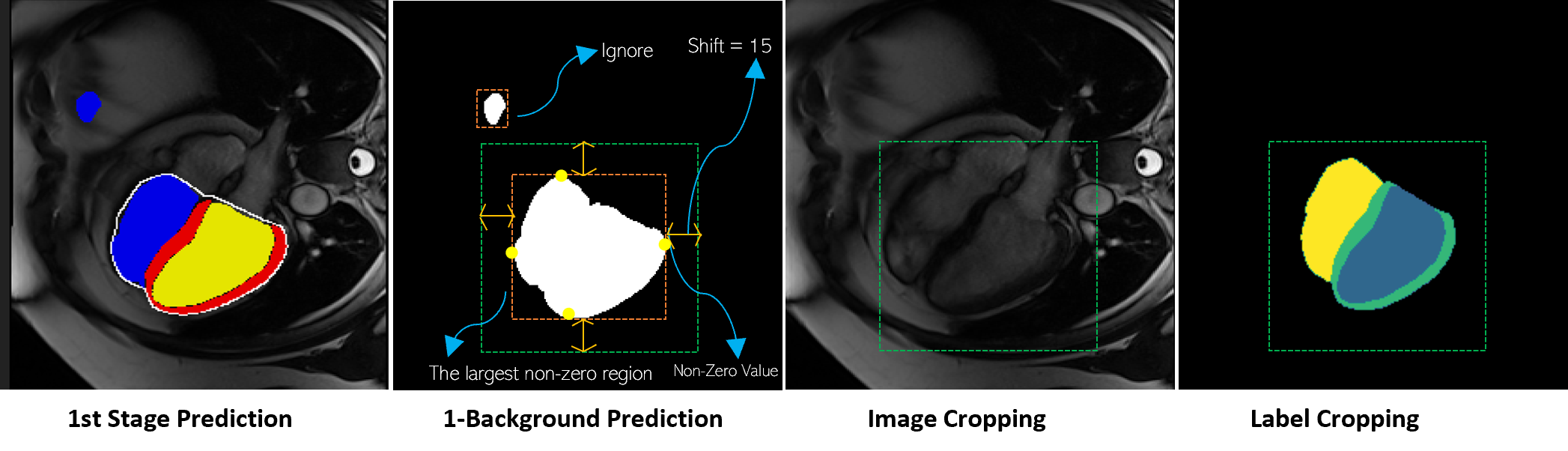}
\caption{Proposed cropping process in CroCNet shows intensity and label images are cropped based on detection of the largest connected region in the inverted background prediction.} 
\label{fig2}
\end{figure}

\textbf{Cropping}: The prediction generated in the first stage utilizes a full-size image, potentially leading to outliers or errors, as illustrated in \figref{fig2}. To address this, we employ background predictions to identify and remove such outliers. The heart region is extracted from both the intensity image and ground truth.
The detection is performed by identifying the largest connected region from element-wise negation of the background predictions. 
To ensure a safety margin that encompasses the entire heart, the bounding box is expanded by \emph{'shift'} of 15 pixels.

\textbf{Specialist Networks}: The specialist networks, lightweight encoder-decoder networks with a similar architecture to E-2AUNet, have three instances (LV-Net, RV-Net, and MYO-Net) with a weight-sharing strategy and cross-efficient additive attention blocks (cross-E-2A), in the encoders. 
The specialist networks work only on the cropped heart regions, having a smaller size compared to the original size image. 
The baseline feature maps start at C1:16 and grow as C2:32 $\rightarrow$C3:64 $\rightarrow$ C4:128 $\rightarrow$ C4:256. Additionally, the specialist networks share weights at all stages of the encoder-decoder. Interactions between features from the three encoders occur through the cross E-2A block. This cross-attention mechanism serves as a soft relative shape prior, aiding the network in rectifying geometric shapes in different areas of interest and thereby contributing to a more precise segmentation, as discussed in the ablation study.

\section{EXPERIMENTS AND RESULTS}\label{sec: EXPERIMENTS AND RESULTS}
\subsection{Dataset Description}
The Multi-Disease, Multi-View and Multi-Center Right Ventricular Segmentation in Cardiac MRI challenge was hosted by MICCAI 2021 to address the segmentation of the RV in cardiac MRI \cite{campello2021multi}. It involves data collected from multiple centres with different views and subjects with various pathologies. This dataset is gathered from three vendors, 
and nine scanners. 
It contains labels for three regions of interest: (i) LV blood pools, (ii) RV blood pools, and (iii) left ventricular MYO. 
The challenge provides 360 labelled instances of two phases of the cardiac cycle, specifically the End-Diastolic and End-Systolic phases.
\begin{table}[h!]
\centering
\caption{Comparison of the results obtained from different methods. 
Methods indicated with a $*$ use multi-view inputs.}\label{tab1}
        \resizebox{0.5\textwidth}{!}{
        \begin{tabular}{|c|c|c|c|c|c|c|c|c|}
            \hline
             \multirow{2}{*}{Methods} &  \multicolumn{4}{|c|}{Dice Score ($\%$) $\uparrow$} & \multicolumn{4}{|c|}{HD (mm) $\downarrow$}\\
             \cline{2-9}
              & LV & RV & Myo & Avg & LV & RV & Myo & Avg\\ 
            \hline
            UNet\cite{ronneberger2015u}    & 87.26 & 88.20 & 79.96 & 85.14 & 13.04 & 08.76 & 12.24 & 11.35\\
            ResUNet\cite{diakogiannis2020resunet}   & 87.61 & 88.41 & 80.12 & 85.38 & 12.72 & 08.39 & 11.28 & 10.80\\
            InfoTrans*\cite{li2021right} &  88.21 & 89.11 & 80.55 & 85.96 & 12.47 & 07.23 & 10.21 & 09.97\\
            TransUNet\cite{chen2021transunet}  & 87.91 & 88.23 & 79.05 & 85.06 & 12.02 & 08.14 & 11.21 & 10.46\\
            MCTrans\cite{ji2021multi}   & 88.42 & 88.19 & 79.47 & 85.36 & 11.78 & 07.65 & 10.76 & 10.06\\
            MCTrans*\cite{ji2021multi}  &  88.81 & 88.61 & 79.94 & 85.79 & 11.52 & 07.02 & 10.07 & 09.54\\
            UTNet\cite{gao2021utnet} &  86.93 & 89.07 & 80.48 & 85.49 & 11.47 & 06.35 & 10.02 & 09.28\\
            UTNet*\cite{gao2021utnet} &  87.36 & {90.42} & 81.02 & 86.27 & 11.13 & {05.91} & {09.81} & {08.95}\\
            TransFusion*\cite{liu2022transfusion}   & {89.78} & {91.52} & {81.79} & {87.70} & {10.25} & {05.12} & {08.69} & {08.02}\\ \hline
            {Proposed-s1(E-2AUNet)} & {93.64} & {91.78} & {84.33} & {89.91} & {3.56} & 5.18 & {3.0}2 & {3.92}\\
             \textbf{Proposed(CroCNet)} & \textbf{95.01} &  \textbf{92.14} &  \textbf{85.53} &  \textbf{90.89} &  \textbf{3.03} &  \textbf{4.10} &  \textbf{2.93} &  \textbf{3.35}\\
             
            \hline
        \end{tabular}}
\end{table} 
\raggedbottom
\subsection{Implementation Details}
CroCNet was implemented in PyTorch using the Adam optimizer and Dice loss. All models are trained for 500 epochs with an initial learning rate of 0.0001, reduced by a factor of 10 after 100 and 200 epochs. We utilized NVidia A100 GPUs with 40GB RAM. Various data augmentation techniques were used, including spatial augmentations (random flip, elastic deformation, random affine) and intensity-based augmentations (random blur, random gamma, random noise).

\subsection{Ablation Studies}
\begin{figure}[ht!]
\centering
\includegraphics[width=6.5cm,height=7cm,trim={0.8cm, 0.5cm, 0.75cm, 0.5cm}]{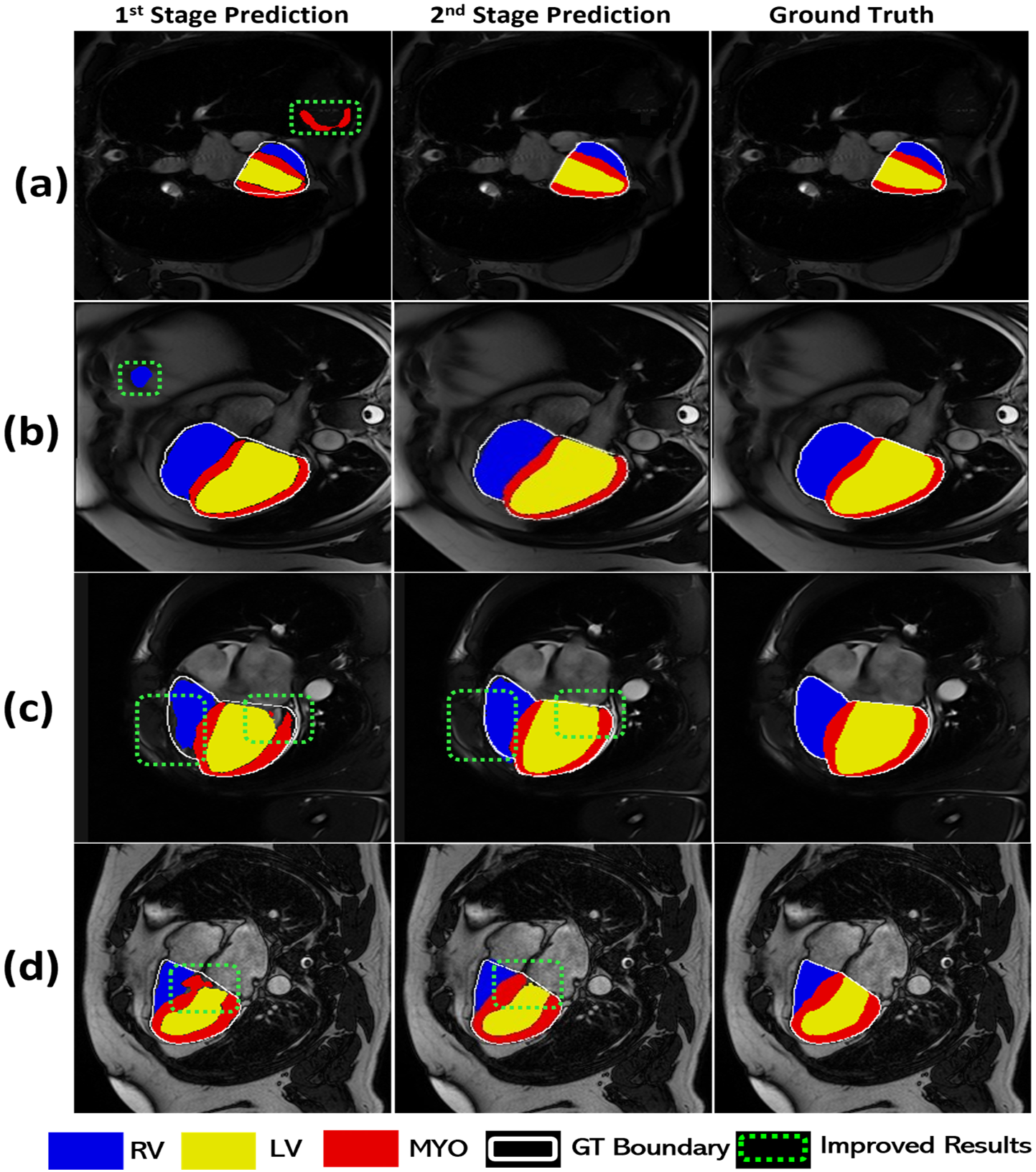}
\caption{Qualitative Results. Row (a) and (b) depict how the cropping process can help to remove erroneous predictions. (c) and (d) showcase the effect of utilizing the Cross-E2A attention with specialist networks.} 
\label{fig3}
\end{figure}
\begin{figure}[ht!]
\centering
\includegraphics[width=5.5cm,height=3.0cm,trim={0.8cm, 0.5cm, 0.75cm, 0.5cm}]{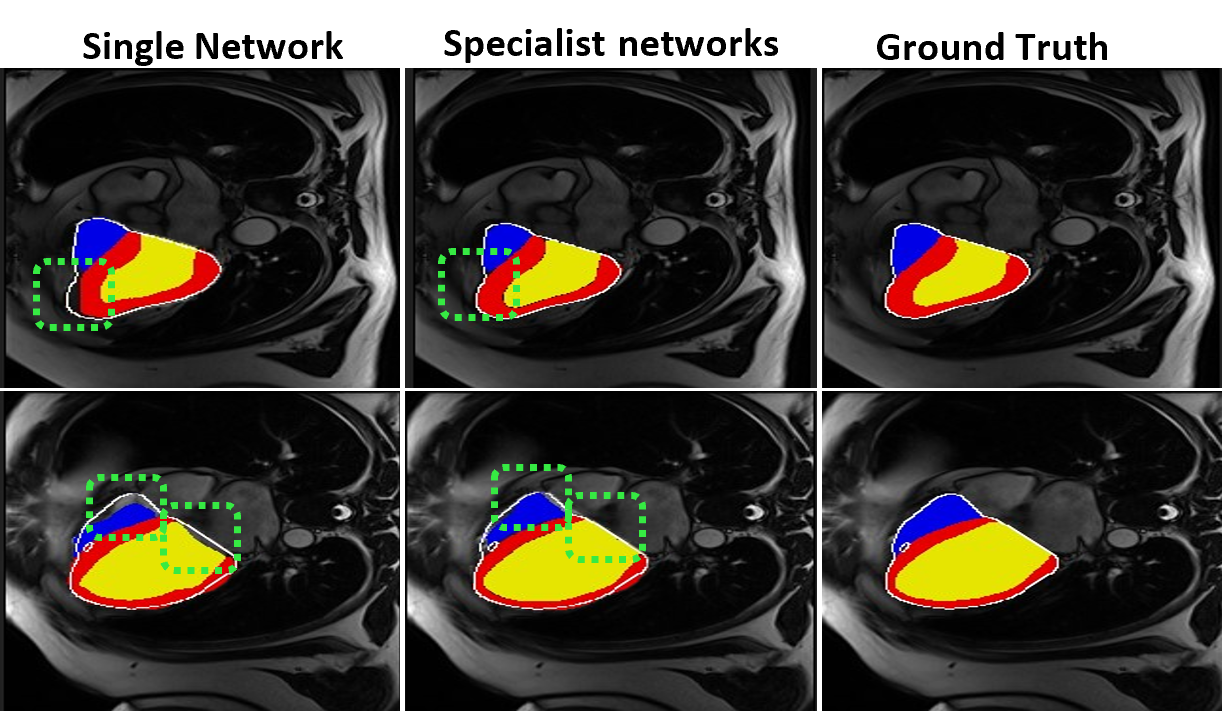}
\caption{Illustration of improved segmentation of boundary regions using specialist network and Cross E-2A.} 
\label{fig4}
\end{figure}

Two ablation studies were conducted to assess the efficacy of the proposed pipeline: (1) excluding the second-stage specialist network and (2) omitting Cross E-2A (i.e. utilizing a single network in the second stage).\\
Our ablation study evaluation, shown in \tabref{tab1}, indicates that the first stage, E-2AUNet, can generate a state-of-the-art result (Proposed-s1). 
\figref{fig3} depicts some problematic cases of stage-1 prediction.
The second stage of CroCNet serves as a refinement step, enhancing the segmentation fidelity from the first stage. The improvement in results is demonstrated in the last two rows in \tabref{tab1}.
In another ablation study, we implemented the second-stage network without the specialist networks, presenting it as a single network without the inclusion of cross-E2A. The results are outlined in \tabref{tab2}. While this strategy remains effective in refining the first stage predictions, there are instances where it might fail to precisely perform segmentation near the boundaries, illustrated in \figref{fig4}.

\begin{table}[hbt!]
\caption{Ablation study for the stage 2 specialist networks. }\label{tab2}
        \resizebox{0.5\textwidth}{!}{
        \centering
        \begin{tabular}{|c|c|c|c|c|c|c|}
            \hline
             \multirow{2}{*}{Stage 2 methods} &  \multicolumn{3}{|c|}{Dice Score ($\%$) $\uparrow$} & \multicolumn{3}{|c|}{HD (mm) $\downarrow$}\\
             \cline{2-7}
              & LV & RV & Myo & LV & RV & Myo\\ 
            \hline
             \makecell{Single network} & {94.44} &  {92.04} &  {84.58} &  {3.10} &  {4.41} &  \textbf{2.86} \\
             \hline
             \makecell{Specialist networks (CroCNet)} & \textbf{95.01} &  \textbf{92.14} &  \textbf{85.53} &  \textbf{3.03} &  \textbf{4.10} &  {2.93} \\
            \hline
        \end{tabular}
        }
\end{table} 
\subsection{Results}
\tabref{tab1} presents a comparative analysis of the proposed method's performance against several state-of-the-art approaches conducted through a five-fold cross-validation split. This comparison includes a variety of architectures, including CNN-based 

\begin{figure}[hbt!]
\centering
\includegraphics[scale=0.25]{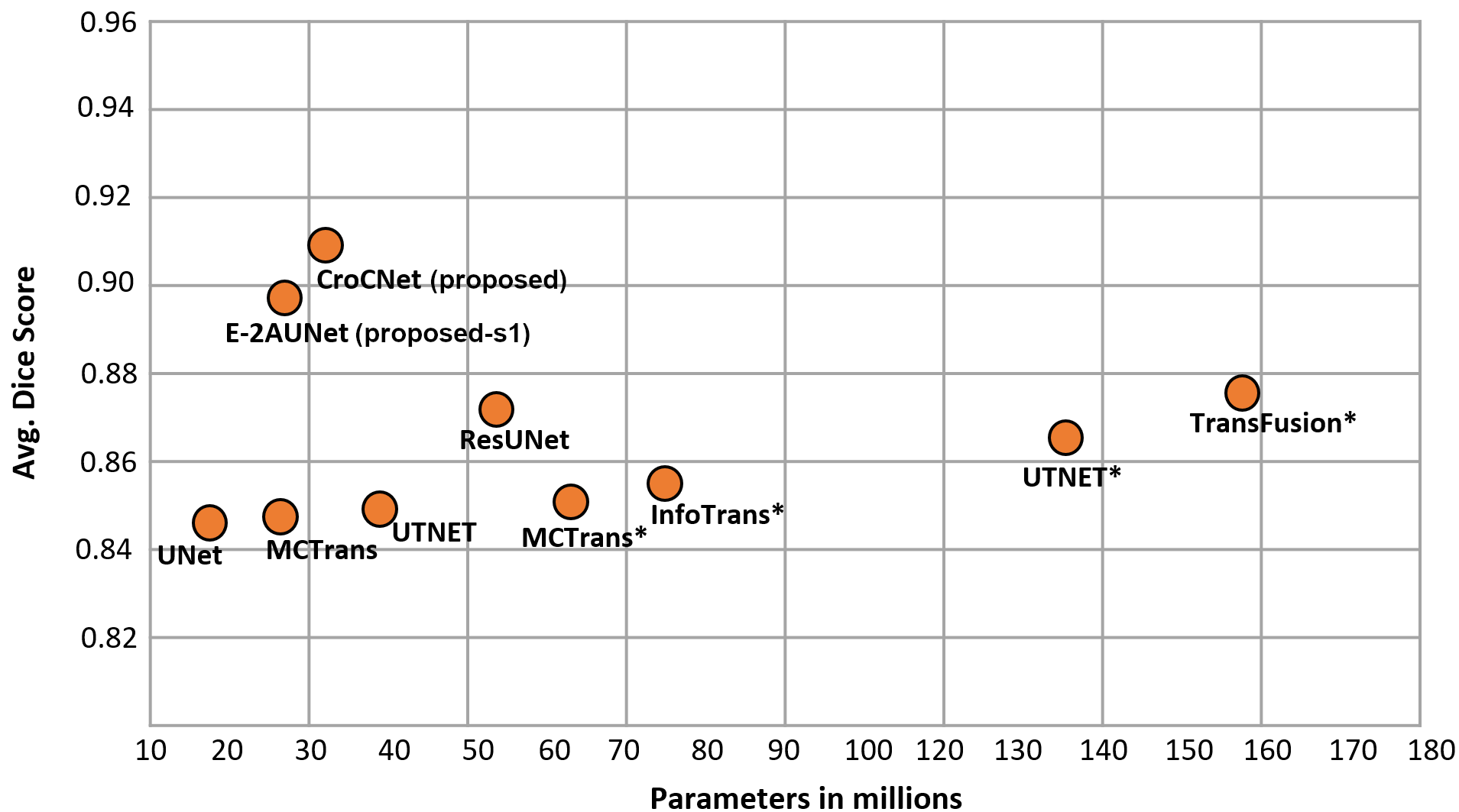}
\caption{Model size versus performance comparison shows average Dice score and number of model parameters.} 
\label{param}
\end{figure}

and hybrid utilizing both CNN and MHSA. \figref{param} depicts the computational complexity of the proposed pipeline compared to existing methods. Results are provided from both stages, i.e. stage 1 (Proposed-s1) and 2 (CroCNet). Notably, the proposed method outperforms existing networks, even at stage 1. The stage 2 network contributes to further refinement, correcting erroneous predictions, as visually depicted in \figref{fig3}.
In \figref{fig3}, the first column shows some erroneous segmentation examples of stage 1 networks. The middle column demonstrates the improvement achieved by the second stage in refining predictions from the first stage. The rightmost column provides the reference ground truth. Additionally, rows (a) and (b) depict how the cropping mechanism aids in removing outliers. Finally, observation from rows (c) and (d) shows that the cross E-2A among the encoder's features of LV-Net, RV-Net, and MYO-Net contributes to further enhance the segmentation maps, particularly addressing disconnected edges in the areas of interest. 

\section{CONCLUSION}\label{sec: CONCLUSION}
This paper introduces CroCNet, a two-stage cardiac image segmentation method, which implements a crop and couple strategy using specialist networks. We proposed E-2AUNet 
which incorporates efficient additive attention in UNet, capturing both local and global contexts with reduced computational complexity than MHSA. In the first stage, 
E-2AUNet was applied to get an initial segmentation. Leveraging this initial segmentation, we proposed a second stage of specialist networks along with a cropping methodology that enabled these networks to focus on only the region of interest. The specialist networks 
utilize a weight-sharing strategy and engage in interactions at all encoder stages through an efficient additive cross-attention. 

\section{Acknowledgments}
This work was funded through the mini-Centre for Doctoral Training in AI-based Cardiac Image Computing provided through the Faculty of Science and Engineering, Queen Mary University of London.  This paper utilised Queen Mary's Andrena HPC facility, supported by QMUL Research-IT Services. Caroline Roney acknowledges funding from a UKRI Future Leaders Fellowship (MR/W004720/1). 
\section{COMPLIANCE WITH ETHICAL STANDARDS}
The licence related to the open-access data ensured no ethical approval was necessary.

\bibliographystyle{IEEEbib}
\bibliography{strings,refs}

\begin{thebibliography}{10}

\bibitem{cohn2003screening}
Jay~N Cohn, Lynn Hoke, Wayne Whitwam, Paul~A Sommers, Anne~L Taylor, Daniel Duprez, Renee Roessler, and Natalia Florea,
\newblock ``Screening for early detection of cardiovascular disease in asymptomatic individuals,''
\newblock {\em Am. Heart J.}, vol. 146, no. 4, pp. 679--685, 2003.

\bibitem{chen2020deep}
Chen Chen, Chen Qin, Huaqi Qiu, Giacomo Tarroni, Jinming Duan, Wenjia Bai, and Daniel Rueckert,
\newblock ``Deep learning for cardiac image segmentation: a review,''
\newblock {\em Front. cardiovasc. med.}, vol. 7, pp. 25, 2020.

\bibitem{ronneberger2015u}
Olaf Ronneberger, Philipp Fischer, and Thomas Brox,
\newblock ``U-net: Convolutional networks for biomedical image segmentation,''
\newblock in {\em MICCAI}. Springer, 2015, pp. 234--241.

\bibitem{de2003adapting}
Marleen de~Bruijne, Bram van Ginneken, Max~A Viergever, and Wiro~J Niessen,
\newblock ``Adapting active shape models for 3d segmentation of tubular structures in medical images,''
\newblock in {\em IPMI 2003, Ambleside, UK,Proceedings 18}. Springer, 2003, pp. 136--147.

\bibitem{kiricsli2010evaluation}
HA~Kiri{\c{s}}li, Michiel Schaap, Stefan Klein, Stella-Lida Papadopoulou, Mara Bonardi, Chin-Hui Chen, Annick~C Weustink, Nico~R Mollet, Evert-Jan Vonken, Rob~J van~der Geest, et~al.,
\newblock ``Evaluation of a multi-atlas based method for segmentation of cardiac cta data: a large-scale, multicenter, and multivendor study,''
\newblock {\em Medical physics}, vol. 37, no. 12, pp. 6279--6291, 2010.

\bibitem{baccouch2023comparative}
Wafa Baccouch, Sameh Oueslati, Basel Solaiman, and Salam Labidi,
\newblock ``A comparative study of cnn and u-net performance for automatic segmentation of medical images: application to cardiac mri,''
\newblock {\em Procedia Comput. Sci.}, vol. 219, pp. 1089--1096, 2023.

\bibitem{deng2021transbridge}
Kaizhong Deng, Yanda Meng, Dongxu Gao, Joshua Bridge, Yaochun Shen, Gregory Lip, Yitian Zhao, and Yalin Zheng,
\newblock ``Transbridge: A lightweight transformer for left ventricle segmentation in echocardiography,''
\newblock in {\em MICCAI Workshops}. Springer, 2021, pp. 63--72.

\bibitem{liu2022transfusion}
Di~Liu, Yunhe Gao, Qilong Zhangli, Ligong Han, Xiaoxiao He, Zhaoyang Xia, Song Wen, Qi~Chang, Zhennan Yan, Mu~Zhou, et~al.,
\newblock ``Transfusion: multi-view divergent fusion for medical image segmentation with transformers,''
\newblock in {\em MICCAI}. Springer, 2022, pp. 485--495.

\bibitem{shaker2023swiftformer}
Abdelrahman Shaker, Muhammad Maaz, Hanoona Rasheed, Salman Khan, Ming-Hsuan Yang, and Fahad~Shahbaz Khan,
\newblock ``Swiftformer: Efficient additive attention for transformer-based real-time mobile vision applications,''
\newblock {\em ICCV}, 2023.

\bibitem{diakogiannis2020resunet}
Foivos~I Diakogiannis, Fran{\c{c}}ois Waldner, Peter Caccetta, and Chen Wu,
\newblock ``Resunet-a: A deep learning framework for semantic segmentation of remotely sensed data,''
\newblock {\em ISPRS Journal of P\&RS}, vol. 162, pp. 94--114, 2020.

\bibitem{zhou2018unet++}
Zongwei Zhou, Md~Mahfuzur Rahman~Siddiquee, Nima Tajbakhsh, and Jianming Liang,
\newblock ``Unet++: A nested u-net architecture for medical image segmentation,''
\newblock in {\em MICCAI Workshops 2018}. Springer, 2018, pp. 3--11.

\bibitem{li2021right}
Lei Li, Wangbin Ding, Liqin Huang, and Xiahai Zhuang,
\newblock ``Right ventricular segmentation from short-and long-axis mris via information transition,''
\newblock in {\em STACOM}. Springer, 2021, pp. 259--267.

\bibitem{carion2020end}
Nicolas Carion, Francisco Massa, Gabriel Synnaeve, Nicolas Usunier, Alexander Kirillov, and Sergey Zagoruyko,
\newblock ``End-to-end object detection with transformers,''
\newblock in {\em ECCV}. Springer, 2020, pp. 213--229.

\bibitem{chen2021transunet}
Jieneng Chen, Yongyi Lu, Qihang Yu, Xiangde Luo, Ehsan Adeli, Yan Wang, Le~Lu, Alan~L Yuille, and Yuyin Zhou,
\newblock ``Transunet: Transformers make strong encoders for medical image segmentation,''
\newblock {\em arXiv preprint arXiv:2102.04306}, 2021.

\bibitem{gao2021utnet}
Yunhe Gao, Mu~Zhou, and Dimitris~N Metaxas,
\newblock ``Utnet: a hybrid transformer architecture for medical image segmentation,''
\newblock in {\em MICCAI}. Springer, 2021, pp. 61--71.

\bibitem{ji2021multi}
Yuanfeng Ji, Ruimao Zhang, Huijie Wang, Zhen Li, Lingyun Wu, Shaoting Zhang, and Ping Luo,
\newblock ``Multi-compound transformer for accurate biomedical image segmentation,''
\newblock in {\em MICCAI}. Springer, 2021, pp. 326--336.

\bibitem{campello2021multi}
Victor~M Campello, Polyxeni Gkontra, Cristian Izquierdo, Carlos Martin-Isla, Alireza Sojoudi, Peter~M Full, Klaus Maier-Hein, Yao Zhang, Zhiqiang He, Jun Ma, et~al.,
\newblock ``Multi-centre, multi-vendor and multi-disease cardiac segmentation: the m\&ms challenge,''
\newblock {\em IEEE Trans on Med Img}, vol. 40, no. 12, pp. 3543--3554, 2021.

\end{thebibliography}

\end{document}